\documentstyle[12pt,fullpage,bbbold,psfig]{article}

\def\slantfrac#1#2{\hbox{$\,^#1\!/_#2$}}

\begin{document}

\title{Driven-disk model for
binaries with precessing donor star.
Three-dimensional simulations}

\author{
Bisikalo D.V.$^1$, Boyarchuk A.A.$^1$,\\
Kuznetsov O.A.$^2$, Chechetkin V.M.$^2$\\[0.3cm]
$^1$ {\it Institute of Astronomy of the Russian Acad. of Sci.,
Moscow}\\
{\sf bisikalo@inasan.rssi.ru; aboyar@inasan.rssi.ru}\\[0.3cm]
$^2$ {\it Keldysh Institute of Applied Mathematics, Moscow}\\
{\sf kuznecov@spp.keldysh.ru; chech@int.keldysh.ru}\\[0.3cm]
}
\date{}
\maketitle

\begin{abstract}
{\bf Abstract}---We present the results of three-dimensional
numerical simulations of mass transfer in semi-detached binary
with a donor star whose rotation vector precesses
around the orbital rotation axis of the binary in the observer's
coordinate frame. The calculations support our previous model of
flow without a `hot spot'. Characteristic features of the flow in
this model, such as the formation of an circumbinary envelope,
the absence of a `hot spot' at the edge of the accretion disk,
and the formation of a shock wave along the edge of the stream,
are also present in the solution for a binary with precessing
donor star.  The parameters of accretion disk and of the
structure of the near-disk regions recur with the precessional
period of the rotation axis of donor star.
\end{abstract}

\section{INTRODUCTION}

Observations of binary systems over the past ten years indicate
that in a number of close binaries, the best known being Her~X-1
(HZ~Her) and SS433, long-period variations are detected on
characteristic timescales substantially longer than the orbital
period. To explain these variations, precession of the
accretion disk in the binary is widely assume (see, e.g., [1--3]
and references therein).

Possible reasons for precession of an  accretion disk have been
analyzed in a number of studies.  As early as in 1972
N.I.Shakura [4] noted that the accretion disk can precess if its
plane does not coincide with the orbital plane of a binary.
Among various mechanisms that can lead to formation of an
accretion disk inclined to the orbital plane, two are usually
considered to be most probable: influence of magnetic field of
the accretor or violation of the symmetry of the donor-star
outflow due to rotation of the star.  The question of
formation an inclined disk under the action of the accretor's
magnetic field remains open: control of the disk orientation
would require a strong magnetic field which, in turn, could
inhibit the formation of the disk itself.

An inclined accretion disk is more likely to form due to a
change of the position of the stream flowing from the inner
Lagrangian point $L_1$. In particular W.Roberts [1] and
A.M.Cherepashchuk [5,6] proposed a scenario in which a minor
asymmetry of the explosion of a supernova (resulting in the
formation of a relativistic object in the binary) can decline
the orbital plane of a binary relative to the rotation axis of
the normal component of the system. In this case, after the
supernova explosion and the formation of a relativistic object,
the rotation axis of the normal component of the system might
become oriented not perpendicular to the orbital plane of the
binary. For systems where mass is transfer from the normal star
to the relativistic object, the disturbance of the symmetry of
the outflow of matter from the donor-star might result in the
formation of an accretion disk not aligned with the orbital
plane. Precession of such a disk could be caused either by
induced precession of the disk itself under the action of the
gravitational attraction of donor-star [7,8], or by the stream
oscillations due to precession of the rotation vector of the
donor-star (the `slaved disk' model, [1,4]), or by some other
mechanism (see, e.g., [9]).

The description of the gas dynamics of mass transfer binaries of
this type calls for the use of three-dimensional models because
the rotation vectors of the donor-star is not perpendicular to
the orbital plane and is engaged in precession.  This means that
this problem cannot be reduced to a two-dimensional one.
Further, there is an additional complexity in describing such
systems connected with the periodic time dependence of the
boundary conditions, that is with the absence of a steady-state
flow of matter in the system, thus necessitating the
consideration of the structure of flow over long periods of time
(few times longer than maximum characteristic period of the
system). Until very recently, these circumstances, together with
insufficient computational resources, made numerical studies of
flow structures in binary systems of this type difficult. a Few
attempts to consider the formation of an accretion disk in such
systems were made in substantially simplified formulation [10,
11].

Here we consider for the first time numerical, self-consistent
solution for the three-dimensional flow structure of mass
transfer in semidetached binary systems in which the donor's
rotation axis precesses. The obtained results support the model
of `driven accretion disk', based on the idea that the
oscillations of the disk relative to the equatorial plane
reflect variations of the stream of matter from the inner
Lagrangian point $L_1$.

\section{THE MODEL}

In [12] we presented a three-dimensional simulation of mass
transfer in semidetached binaries with rotation of the
donor-star. We investigated both synchronous (the rotational
period of the donor-star equals to the orbital period of the
system $P_\star=P_{orb}$) and asynchronous ($P_\star\ne
P_{orb}$) rotation of the donor-star [12]. For the case of
asynchronous rotation both axially aligned rotation, when the
rotation vector of the donor-star is perpendicular to the
orbital plane of a binary, and non-aligned, when this vector is
inclined relative to the orbital plane were considered. It was
also assumed in [12] that the case of asynchronous non-aligned
rotation implies that the rotation vector of the donor-star is
in a rest in the laboratory coordinate system (i.e., relative to
observer).

The existence of observational evidence for precession of
rotational vector of the donor-star requires extension of the
model.  The model we use here assumes that the
rotation vector of the donor-star precesses (in a laboratory
coordinate system) about its mean position, which coincides with
the vector of orbital rotation of the binary. Note that although
this model has a formally more general character (rate of
precession of the rotation vector in the laboratory coordinate
system ${\tilde\Omega}_{lab}\neq0$) than the model considered in
[12] for the case of asynchronous non-aligned rotation (rate
of precession of the rotation vector in the laboratory
coordinate system ${\tilde\Omega}_{lab}=0$), these models differ
significantly in the laboratory coordinate system only. In a
rotating coordinate system (adopted for our calculations), we
can expect that there will be qualitatively similar solutions
for both models because in this coordinate system the rotation
vector of the donor-star in model of [12] also undergoes
precessional motion, i.e. for both models
${\tilde\Omega}_{rot}\neq0$. The different rates of precession
in the models should lead only to quantitatively difference of
the solutions leaving the general features of the flow structure
in a binary unchanged.

To study in detail whether accretion disk will follow the
`driven disk' model, we conducted the calculations over a time
exceeding the period of long-period variations of the system;
this made it possible to obtain established solutions both in a
laboratory and in a rotating coordinate system.  To be able to
compare the solutions with the results obtained in [12] we
adopted the same parameters of binary system: the primary
(filling its Roche lobe) parameters are $M_1=0.28 M_\odot$,
$T_1=10^4$ K; the secondary (compact objects) parameters are
$R_2=0.013R_\odot$, $M_2=1.4 M_\odot$; parameters of binary
system $P_{orb}=5^{\mbox{h}}\!\!.56$; $A=1.97 R_\odot$. We
assumed that the rotation velocity of the donor-star in the
laboratory coordinate system exceeds twice the angular velocity
of the system. The angle between the rotation vector of the
donor star ${\bmath\Omega}_\star$ and the `$z$'-axis was assumed
to be $\vartheta=15^\circ$ (in the laboratory coordinate
system).\footnote{According to the vector transformation rule
${\bmath\Omega}_{rot}={\bmath\Omega}_{lab}-{\bmath\Omega}$ the
inclination angle of this vector in the rotating coordinate
system is $\vartheta^*=30^\circ$}. The rate of precession of
rotation vector of the donor-star was equal to
${\tilde\Omega}_{lab}=\slantfrac{1}{6}\Omega$ in the laboratory
coordinate system or
${\tilde\Omega}_{rot}=-\slantfrac{5}{6}\Omega$ -- in the
rotating coordinate system (the minus sign indicates that the
precession of the rotation vector is retrograde with respect to
the orbital rotation).

Following the procedure of [12], we determined the boundary
conditions on the surface of this star by solving the Riemann
problem between gas parameters ($\rho_1,~{\bmath v}_1,~p_1$) on
the stellar surface and the parameters in the computation
gridpoint closest to the given point on the star surface.  The
asynchronism of rotation of the donor-star, as well as the
precession of the vector of its rotation were taken into account
when setting the boundary conditions for the gas velocity vector
on the Roche lobe of the donor-star. All other parameters of the
mathematical model including the details of numerical
realization were taken the same as in [12], namely:

\begin{itemize}

\item We adopted the computational domain as a parallepipedon
$(-A\ldots 2A)\times(-A\ldots A)\times(-A\ldots A)$;

\item A non-uniform (more fine in the zone near accretor)
finite-difference grid with $91\times81\times55$ gridpoints was
used;

\item The shape of the donor-star was determined, taking into
account the asynchronicity of its rotation.

\end{itemize}

\renewcommand{\thefigure}{1}
\begin{figure}[t]
\centerline{\hbox{\psfig{figure=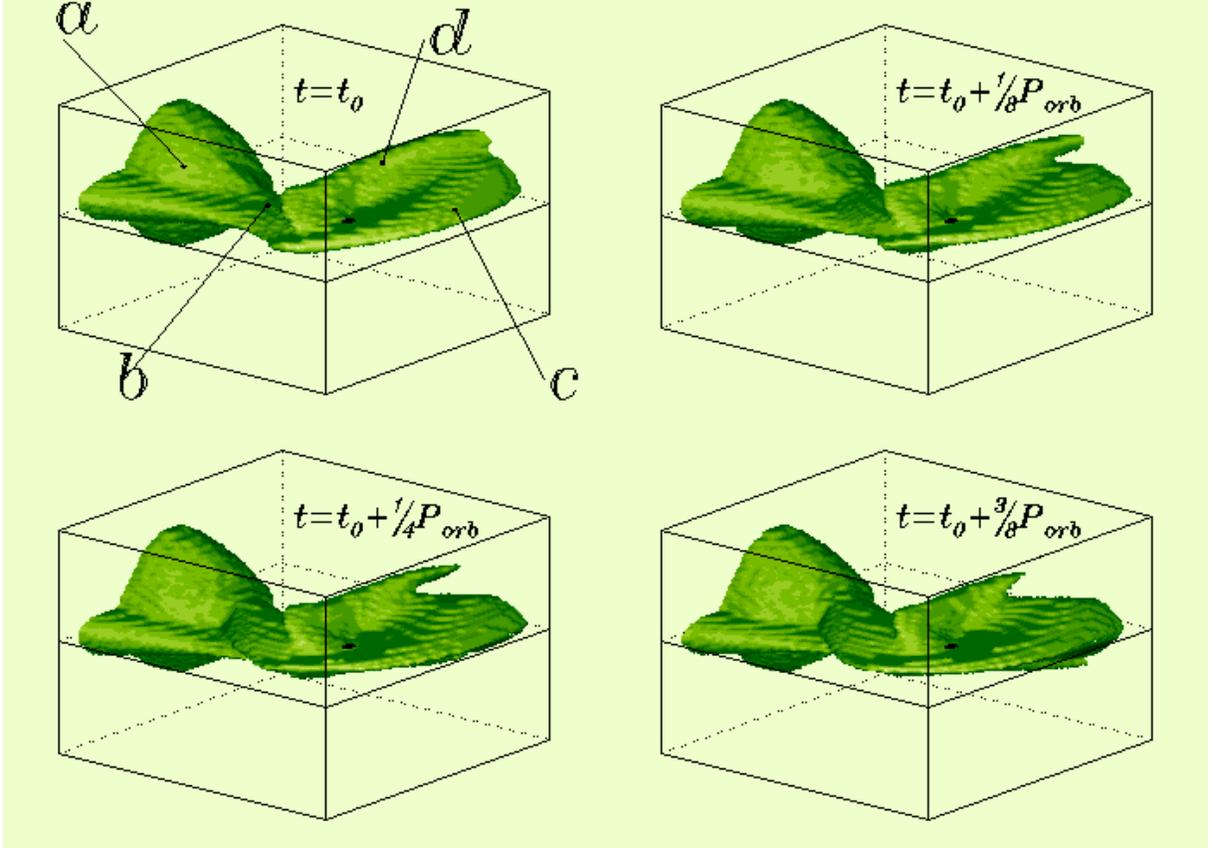,width=16cm}}}
\caption{\small
Iso-surfaces of density at level $\rho=0.002\rho_{L_1}$ for
first four moments of time covering the full period of boundary
conditions variations in the rotating coordinate system. Filled
circle is the accretor. On the first panel corresponding to
moment of time $t=t_0$, the basic features of the flow pattern
are marked by $a$, $b$, $c$, $d$. The same features are also
presented on all other panels: $a$ -- the donor-star surface;
$b$ -- the gas stream from $L_1$; $c$ -- the flow region near
the accretor including the accretion disk; $d$ -- the
gas `clouds' located in front of the forward edge of the stream
outside of the disk formation region.}
\end{figure}

\renewcommand{\thefigure}{1 {\small \it (continued)}}
\begin{figure}[p]
\centerline{\hbox{\psfig{figure=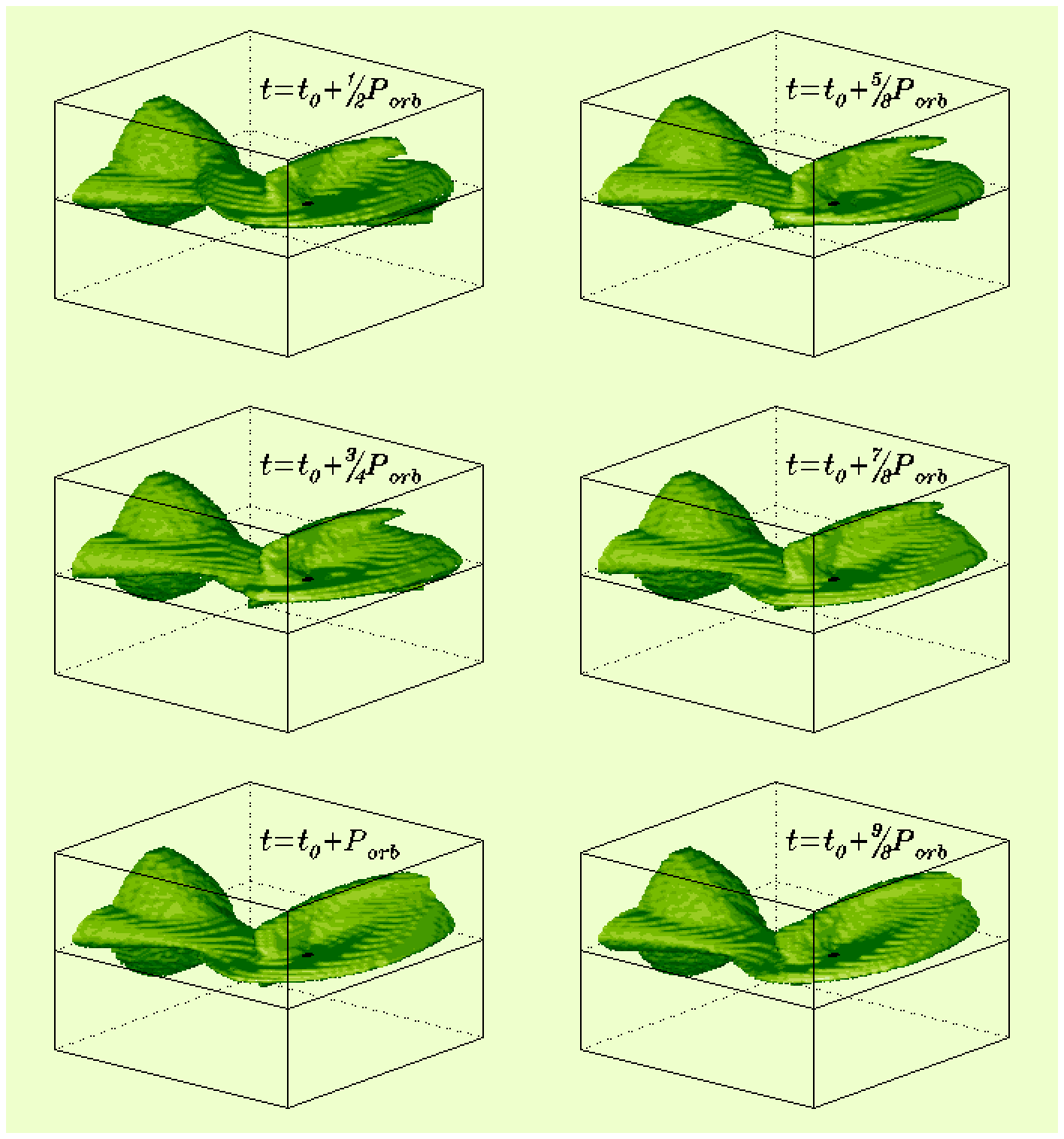,width=16cm}}}
\caption{\small
Iso-surfaces of density at level $\rho=0.002\rho_{L_1}$ for
last six moments of time covering the full period of boundary
conditions variations in the rotating coordinate system.}
\end{figure}

\section{RESULTS OF CALCULATIONS}

The results of calculations fully confirmed our assumption that,
given a rotating coordinate system, the qualitative features of
the flow structure are independent of the rate of precession of
the rotation vector of the donor-star. Fig.~1 depicts the
iso-surfaces of density at the level $\rho=0.002\rho_{L_1}$ for
10 moments of time which cover a full period of variation of the
boundary conditions in a rotating coordinate system. Note that,
in line with the assumed parameters of the system, the period of
precession of the rotation vector of the donor-star in a
rotating coordinate system ${\tilde P}_{rot}$ is equal to
$\slantfrac{6}{5}$ of the orbital period $P_{orb}$. Owing to
this the results are given in Fig.~1 over the interval
$\slantfrac{6}{5}P_{orb}$.

Analysis of the results presented in Fig.~1 shows that the
behavior of the disk $c$ and of the near-disk matter $d$
reflects the variations of the stream of matter $a$ flowing from
$L_1$ (i.e. boundary conditions on the surface of the donor-star
$a$).  This points to the realization of the `slaved disk' model
in the calculations. The results show also that the unique
morphology of the flow, as well as the effect of circumbinary
envelope gas lead to a shock-free interaction between the stream
and the outer edge of the accretion disk and, as a consequence,
to the absence of a `hot spot' in the disk. In the model
discussed here the zone of increased energy release is located
beyond the accretion disk -- at the region where the gas of
circumbinary envelope interacts with the stream and where the
extended shock wave is formed. Similarly to the case of
non-aligned asynchronous rotation described in [12], the
interaction between the rarefied gas of the circumbinary
envelope and the stream of matter results in the formation of
gaseous `clouds' located ahead of the front edge of the stream,
beyond the zone of disk formation (marker $d$ in Fig.~1).  The
period of variation of these gaseous formations -- `clouds' --
reflects the period of variation of the boundary conditions at
the mass-losing star.

In the laboratory coordinate system the characteristic timescale
of the long-period variation (precession of the rotation vector
of the donor-star with period ${\tilde P}_{lab}$) is equal to
$6P_{orb}$. The calculations were made over period of time
exceeding ${\tilde P}_{lab}$ starting from the time when the
flow was established.  As expected, the comparison of results of
calculations in the laboratory coordinate system confirms the
periodic character of solution with the period ${\tilde
P}_{lab}$.  This is illustrated in Fig.~2 and~3 where
iso-surfaces of density at the level
$\rho=0.002\rho_{L_1}$ are shown in the laboratory
coordinate system. These plots show the 3D view of density
distribution in the vicinity of $L_1$ and in the near-disk
region, respectively, for six moments of time covering the
period of precession in the laboratory coordinate system.
Analysis of Fig.~2 indicates that in accordance with the
variations of boundary conditions, the position of the stream
flowing from $L_1$ varies with respect to the equatorial plane,
and, after a time ${\tilde P}_{lab}$, returns to its initial
position. Figure~3 shows that the accretion disk $c$ and
surrounding gaseous formation (clouds $d$ and $d'$) follow the
position of the stream (see Fig.~2), and their shape varies with
the same period.

\renewcommand{\thefigure}{2}
\begin{figure}[p]
\centerline{\hbox{\psfig{figure=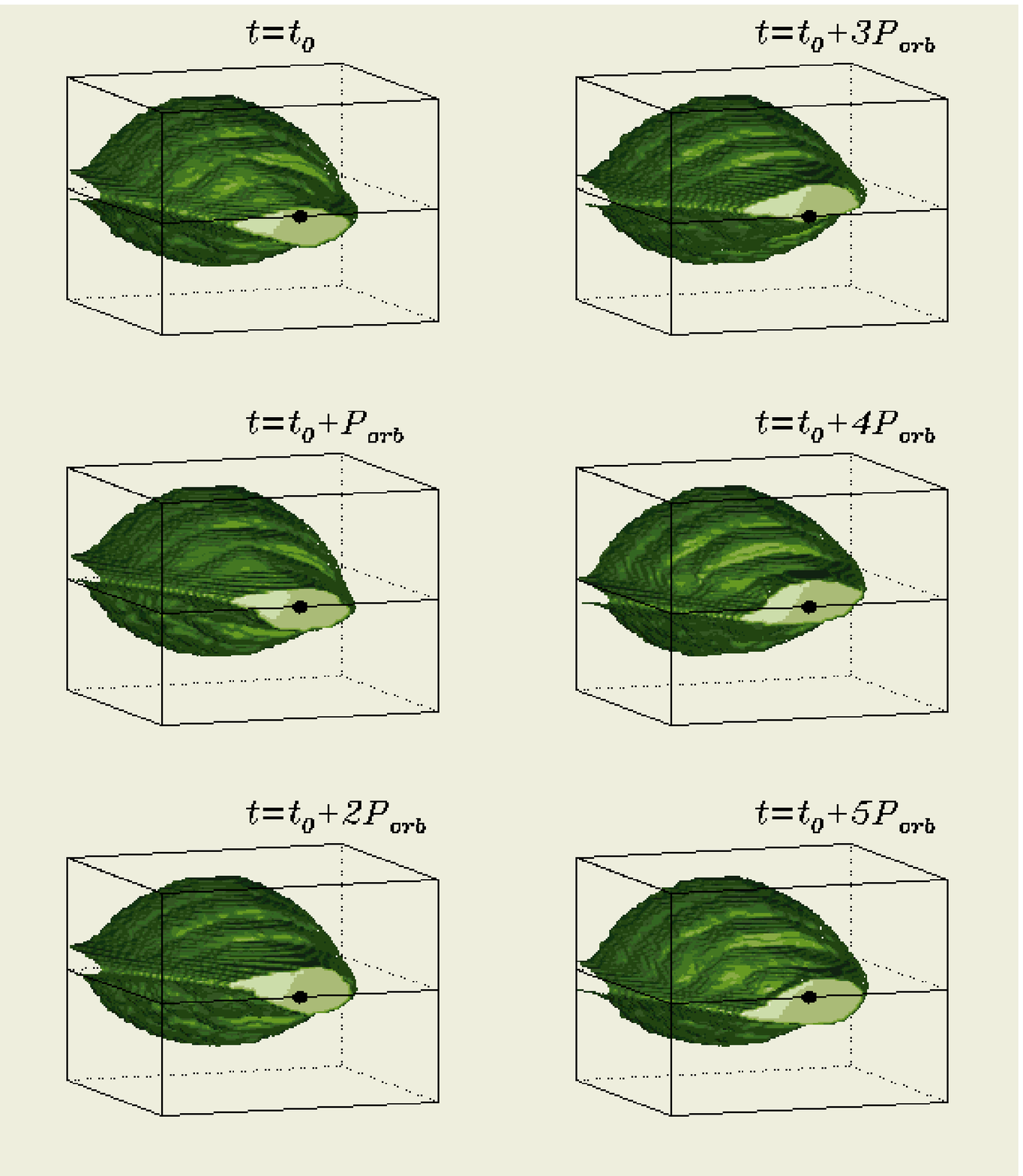,width=16cm}}}
\caption{\small
Iso-surfaces of density at level $\rho=0.002\rho_{L_1}$ in the
vicinity of the inner Lagrangian point for six moments
of time covering the precession period in the laboratory
coordinate system. The iso-surface is cut by plane `$yz$' at a
distance of $0.066A$ from $L_1$. Filled circle is the projection
of the inner Lagrangian point onto the `$yz$' cross-section.}
\end{figure}

\renewcommand{\thefigure}{3}
\begin{figure}[p]
\centerline{\hbox{\psfig{figure=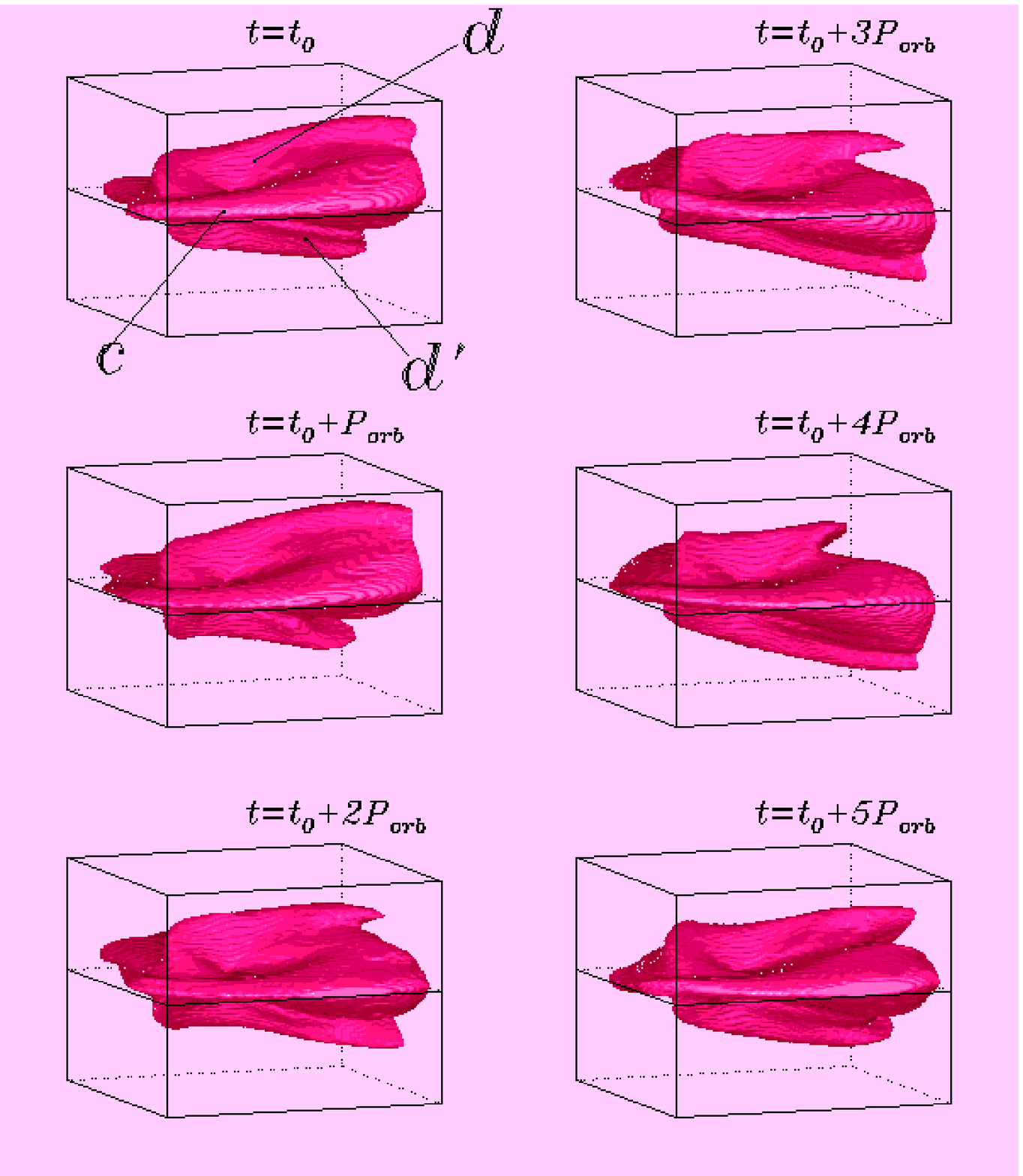,width=16cm}}}
\caption{\small
Iso-surfaces of density at level $\rho=0.002\rho_{L_1}$ in the
vicinity of the accretor for six moments of time, covering the
precession period in the laboratory coordinate system.  On the
first panel corresponding to moment of time $t=t_0$, the basic
features of the flow pattern are marked by the same characters
as in Fig.~1: $c$ -- the flow region near the accretor including
the accretion disk; $d$ and $d'$ -- the gas `clouds' located in
front of the forward edge of the stream outside of the disk
formation region.}
\end{figure}

The presence of two characteristic periods -- the precessional
period of the rotation vector of the donor star in the
laboratory frame ${\tilde P}_{lab}$ and orbital period of the
system $P_{orb}$ -- can explain the appearance of other periods
of observation evidences, apart from ${\tilde P}_{lab}$.  In
particular, the interaction of ${\tilde P}_{lab}$ with $P_{orb}$
leads to formation of short-period beating oscillations. For
example, as concerns the system SS433
($P_{orb}=13^{\mbox{d}}\!\!.1$,
${\tilde P}_{rot}=12^{\mbox{d}}\!\!.1$),
in additions to the period
${\tilde P}_{lab}=162^{\mbox{d}}\!\!.5$, there should exist
variations with the period of $6^{\mbox{d}}\!\!.28$. These
short-period oscillations, as well as the following beating
harmonic with period $5^{\mbox{d}}\!\!.83$ are observed in SS433
(see, e.g., [13,14] and references therein).

\section{CONCLUSIONS}

Our three-dimensional simulations of mass transfer in
semidetached binaries with precession of rotation vector of
donor star reveal the `driven' nature of forming accretion disk.
For typical parameters of numerical viscosity adopted for the
numerical model ($\alpha \sim 0.1\div0.5$ in terms of
$\alpha$-disk), the change of the flow pattern, the parameters
of accretion disk, and parameters of the near-disk regions
reflect the variations of boundary conditions on the donor star.
In turn, the periodicity of boundary conditions is determined by
the precessional velocity. Analysis of the flow pattern
indicates that the basic features of the solution are
qualitatively similar to those for calculations previously
obtained for the cases of synchronous, aligned asynchronous, and
non-aligned asynchronous rotation of the donor-star [12], and,
in turn, indicates to the universal character of the model
without a `hot spot' proposed in [15--18].

The `driven' character of the solution implies that the emission
properties of the accretion disk and intercomponent gaseous
structures recur with the precessional period of the rotation
axis of donor-star. In binary systems where observed long-period
variations can be explained by the precession of the donor-star
the periodicity of the solution obtained here can be used to
interpret the observational data.

\section*{ACKNOWLEDGMENTS}

   This work was supported by the Russian Foundation for Basic
Research (grant 99-02-17619), the INTAS Foundation
(grant 93-93-EXT), and Russian Federation Presedential Grant N
99-15-96022.

\section*{REFERENCES}

\begin{enumerate}

\item Roberts, W.J. 1974, ApJ, 187, 575

\item Cherepashchuk, A.M. 1988, Itogi Nauki i Techniki.
Astronomiya, 38, 60 (in Russian)

\item Margon, B. 1984, ARA\&A, 22, 507

\item Shakura, N.I. 1972, Astron. Zh., 49, 921
(Sov. Astron., 16, 756)

\item Cherepashchuk, A.M. 1981, Pis'ma Astron. Zh., 7, 201
(Sov. Astron. Lett., 7, 111)

\item Cherepashchuk, A.M. 1981, Pis'ma Astron. Zh., 7, 726
(Sov. Astron. Lett., 7, 401)

\item Katz, J.I. 1973, Nature Phys. Sci., 246, 87

\item Petterson, J.A. 1975, ApJ, 201, L61

\item Shakura, N.I., Prokhorov, M.E., Postnov, K.A. \&
Ketsaris, N.A. 1999, A\&A, 348, 917

\item Belvedere, G., Lanzafame, G., \& Molteni, D.
1993, A\&A, 280, 525

\item Lanzafame, G., Belvedere, G., \& Molteni, D. 1994, MNRAS,
267, 312

\item Bisikalo, D.V., Boyarchuk, A.A., Kuznetsov, O.A., \&
Chechetkin, V.M. 1999, Astron. Zh., 76, 270
(Astron. Reports, 43, 229; preprint astro-ph/9812484)

\item Katz, J.I., Anderson, S.F., Margon, B., \& Grandi, S.A.
1982, ApJ, 260, 780

\item Margon, B., \& Anderson, S.F. 1989, ApJ, 347, 448

\item Bisikalo, D.V., Boyarchuk, A.A., Kuznetsov, O.A., \&
Chechetkin, V.M. 1997, Astron. Zh., 74, 880 (Astron. Reports,
41, 786; preprint astro-ph/9802004)

\item Bisikalo, D.V., Boyarchuk, A.A., Kuznetsov, O.A., \&
Chechetkin, V.M. 1997, Astron. Zh., 74, 889 (Astron. Reports,
41, 794; preprint astro-ph/9802039)

\item Bisikalo, D.V., Boyarchuk, A.A., Chechetkin, V.M.,
Molteni, D., \& Kuznetsov, O.A. 1998, MNRAS, 300, 39

\item Bisikalo, D.V., Boyarchuk, A.A., Kuznetsov, O.A., \&
Chechetkin, V.M. 1998, Astron. Zh., 75, 706
(Astron. Reports, 42, 621; preprint astro-ph/9806013)

\end{enumerate}

\end{document}